\documentclass[aps,prd,twocolumn,amssymb,nofootinbib]{revtex4}
\usepackage{graphicx}
\usepackage{amsmath}
\usepackage{amssymb}
\usepackage{bm}

\def\mpl{M_{\rm p}}
  \def\be{\begin{equation}}
  \def\beq{\begin{equation}}
\def\ee{\end{equation}}
\def\eeq{\end{equation}}
\def\bea{\begin{eqnarray}}
\def\eea{\end{eqnarray}}

\def\mpl{M_{Pl}}

\def\sq2{\sqrt{2}}

\def\ph{\varphi}

\def\m4{m^4(\ph)}
\def\mn2{m_n^2}

\def\v5{V^{(5)}}
\newcommand{\half}{{1\over 2}}
\newcommand{\nn}{\nonumber}

\begin{document}

\title{Instabilities of Spherical Solutions with Multiple Galileons and $SO(N)$ Symmetry}
\author{Melinda Andrews}
\email[mgildner@sas.upenn.edu]{}
\author{Kurt Hinterbichler}
\email[kurthi@physics.upenn.edu]{}
\author{Justin Khoury}
\email[jkhoury@sas.upenn.edu]{}
\author{Mark Trodden}
\email[trodden@physics.upenn.edu]{}
\affiliation{Center for Particle Cosmology, Department of Physics and Astronomy, University of Pennsylvania,  209 S 33rd Street, Philadelphia PA 19104-6396 USA}
\date{\today}

\begin{abstract}
\noindent  The 4-dimensional effective theory arising from an induced gravity action for a co-dimension greater than one brane consists of multiple galileon fields $\pi^I$, $I=1,\ldots,N$, invariant under separate Galilean transformations for each scalar, and under an internal $SO(N)$ symmetry.  We study the 
viability of such models by examining spherically symmetric solutions. We find that for general, non-derivative couplings to matter invariant
under the internal symmetry, such solutions exist and exhibit a Vainshtein screening effect. By studying perturbations about such solutions, we find both an inevitable gradient instability and fluctuations propagating at superluminal speeds. These findings suggest that more general, derivative couplings to matter are required for the viability of $SO(N)$ galileon theories.
\end{abstract}

\maketitle

\section{Introduction}
There has been much recent interest in theories of gravity arising from scenarios with extra spatial dimensions. Many examples of these are based on the Dvali-Gabadadze-Porrati (DGP) model~\cite{Dvali:2000hr,DGP2} --- a $4+1$ dimensional theory with action consisting simply of separate Einstein-Hilbert terms in the bulk and
on a codimension-1 brane, to which standard model particles are also confined. The model results in a 4D gravitational force law at sufficiently small scales, which transitions to a 5D gravitational force law at a crossover length scale $r_c\sim M_{\rm Pl}^2/M_5^3$, determined by the $5$D and $4$D gravitational couplings $M_5$ and $M_{\rm Pl}$ respectively.  To yield interesting cosmological dynamics, this crossover scale is usually chosen to be of order the horizon size. 

Much of the phenomenology of the DGP model is captured by its decoupling limit $M_{\rm Pl}$, $M_5\,\rightarrow \infty$ with the strong-coupling scale $\Lambda_5 \sim { M_5^2 / M_{\rm Pl} }$ kept fixed~\cite{Luty:2003vm,Nicolis:2004qq}.  In this limit, the difference between DGP gravity and general relativity is encoded in the behavior of a scalar degree of freedom, $\pi$. The dynamics of this scalar are invariant under internal galilean transformations $\pi \rightarrow \pi + c + b_\mu x^\mu$, with $c$ a constant and $b_{\mu}$ a constant vector. This symmetry proves to be extremely restrictive, with a leading order self-interaction term which is a higher-derivative coupling cubic in $\pi$, and yet which yields second order equations of motion. Higher order couplings with these properties were derived independently of the DGP model~\cite{Nicolis:2008in,covariant,Deffayet:2009mn,Nicolis:2009qm} and dubbed ``galileons." See~\cite{nathan,galileoncosmo1,galileoncosmo2,galileoncosmo3,nbody,vpec,genesis,DeFelice:2010pv,DeFelice:2010nf} for cosmological studies of galileon theories.

It is natural to explore induced gravity models in co-dimension greater than one~\cite{sergei,gigashif,massimo1,cascade1,cascade2,nemanjacascade,cascadeothers1,cascade3,cascade4}, and recently multi-galileon actions arising in the relevant $4$-dimensional decoupling limit  have been derived~\cite{Hinterbichler:2010xn,Deffayet:2010zh,Padilla:2010ir,Padilla:2010de}.
The theories studied in \cite{Hinterbichler:2010xn} are invariant under individual galilean transformations of the $\pi$ fields, and also under an internal $SO(N)$ symmetry rotating the fields into one another, thus forbidding the existence of terms containing an odd number of $\pi$ fields, in contrast to the co-dimension one DGP case.

In this paper we explore the nature of spherically symmetric solutions in theories with an $SO(N)$ internal symmetry among the galileon fields, and 
couplings to matter that respect this symmetry.
Spherical solutions for a more general bi-galileon action were discussed in~\cite{Padilla:2010tj}, for the specific case of a linear coupling $\sim \pi T$ to matter, where $T$ is the trace of the matter energy momentum tensor.  This form of coupling arises from decoupling limits of DGP-like theories, because $\pi$ arises through a conformal mixing with the graviton.  However, while this coupling is therefore the natural form to consider in the case of a single galileon field, it breaks the new internal symmetry satisfied by 
multiple galileons (and breaks the galilean symmetry if the matter is dynamical). We instead study general non-derivative couplings to matter
fields which respect the $SO(N)$ internal symmetry.  

At the background level, our solution can always be rotated to lie along a single field direction, say $\pi_1$, while the other field variables remain trivial, thus exhibiting
spontaneous symmetry breaking. The solution exhibits
Vainshtein screening~\cite{vainshtein,ddgv}, characteristic of galileon theories: we find $\pi_1 \sim r$ sufficiently close to the source, whereas $\pi_1\sim 1/r$ far away, with the crossover scale determined by a combination of the galileon self-interaction scale and the coupling to the source. However, when we turn to the stability of spherically symmetric solutions under small perturbations, we find that, sufficiently close to the source, perturbations in $\pi_1$ suffer from gradient instabilities along the angular directions. Morever, they propagate superluminally both along the radial and angular directions (in the regime that angular perturbations are stable). Perturbations in the remaining $N-1$ galileon fields are stable but propagate superluminally in the radial direction. 

The gradient instability and superluminal propagation found here for the $\pi_1$ field are multi-field generalizations of single galileon
instabilities~\cite{Nicolis:2008in}. Our findings thus present significant hurdles for $SO(N)$ galileon models with non-derivative matter coupling. One of the main lessons to be drawn
is that more general matter couplings, including derivative interactions, are necessary for the phenomenological viability of $SO(N)$ multi-galileon theories. For instance,
the coupling $\sim\partial_{\mu}\pi^I\partial_{\nu}\pi_I T^{\mu\nu}$ naturally arises from brane-world constructions~\cite{deRham:2010eu,Hinterbichler:2010xn} and maintains both the galilean and the internal rotation symmetries.

\section{The model\label{s:model}}

In co-dimension $N$, the $4$-dimensional effective theory contains $N$ fields  $\pi^I$, $I=1\cdots N$, representing the $N$ brane-bending modes of the
full $4+N$-dimensional theory.  The extended symmetry of the vacuum lagrangian is 
\be
\delta \pi^I=\omega^I_{\ \mu}x^\mu+\epsilon^I+\omega^I_{\ J}\pi^J \ ,
\label{multisymmetry}
\ee
where $\omega^I_{\ \mu}$, $\epsilon^I$ and $\omega^I_{\ J}$ are constant transformation parameters. (See \cite{Hinterbichler:2010xn} for the geometric setup and origin of this symmetry).  This transformation consists of a galilean invariance acting on each of the $\pi^I$ fields, and an $SO(N)$ rotation symmetry under which $\pi_I$ transforms as a vector. The unique four dimensional lagrangian density respecting this is~\cite{Padilla:2010ir,Hinterbichler:2010xn}
\bea 
\mathcal{L}_\pi = &&- \ \half\partial_\mu\pi^I\partial^\mu\pi_I  \nn \\  &&-\lambda\left[ \partial_\mu\pi^I\partial_\nu\pi^J\left(\partial_\lambda\partial^\mu\pi_J\partial^\lambda\partial^\nu\pi_I-\partial^\mu\partial^\nu\pi_I\square\pi_J \right)\right], \nn \\ 
\eea
where $\lambda$ is a coupling with dimension [mass]$^{-6}$, containing the strong interaction mass scale.  The $I,J$ indices are raised and lowered with $\delta_{IJ}$.  

It remains to couple this theory to matter.  The natural coupling we might consider, the lowest dimension coupling that preserves the galilean and internal rotation symmetries, is $\sim\partial_{\mu}\pi^I\partial_{\nu}\pi_I T^{\mu\nu}$.  This is the coupling that naturally arises from brane matter in the construction of~\cite{deRham:2010eu,Hinterbichler:2010xn}. However, for static non-relativistic sources $T^{\mu\nu}\sim \rho \delta^\mu_0\delta^\nu_0$, and since $\partial_0\pi=0$ for static solutions there are no nontrivial spherically symmetric solutions with this coupling.  

Linear couplings ${\cal L}_{\rm linear}\sim\pi T$ arise naturally from DGP-like setups, since the $\pi$'s conformally mix with the graviton.  These lead to spherical solutions~\cite{Padilla:2010tj}, but break the $SO(N)$ internal symmetry.

We therefore do not consider these couplings further, and instead concentrate on the most general non-derivative coupling that preserves the $SO(N)$ symmetry.
\be
\mathcal{L}_{\rm coupling} = {T\over 2 } P(\pi^2)\ ,
\label{couplings}
\ee
where $P$ is an arbitrary function of the invariant $\pi^2\equiv \pi^I\pi_I$. 

\section{Spherically symmetric solutions\label{s:solution}}
Our focus is on the existence and viability of spherically symmetric solutions sourced by a delta function mass distribution\footnote{Note that stable, non-trivial solutions without a source do not exist \cite{Endlich:2010zj}.}
\be
T = -M \delta^3(r) \ .
\ee 
The equations of motion, including the coupling~(\ref{couplings}), are
\bea 
\square \pi^I \nn  
&-&\lambda \left[\square\pi^I \left(\partial_\mu\partial_\nu\pi_J\partial^\mu\partial^\nu\pi^J- \square\pi^J\square\pi_J\right)\right. \\ \nn
&+&\left.2\partial_\mu\partial_\nu\pi^I \left(  \partial^\mu\partial^\nu\pi_J\square\pi^J -\partial^\mu\partial_\lambda\pi_J\partial^\nu\partial^\lambda\pi^J\right)\right] \\ && =MP'\left(\pi^2(0)\right)\pi^I(0)\delta^3(\vec r)\ ,
\eea
where $P'(X) \equiv {\rm d}P/{\rm d}X$. Restricting to spherically symmetric configurations $\pi^I(r)$, this reduces to
\be {1\over r^2}{{\rm d}\over {\rm d}r}\left[r^3\left(y^I+2\lambda y^I y^2\right)\right] = MP'\left(\pi^2(0)\right)\pi^I(0)\delta^3(\vec r)\ ,\ee
where 
\be y^I\equiv \frac{1}{r}{{\rm d}\pi^I \over {\rm d}r}\ ,\ee
and $y^2\equiv y^Iy_I$.   Note that, due to the shift symmetry of the lagrangian, the equations of motion of galileon fields always take the form of a total derivative.  Thus we can integrate once to obtain the equations of motion
\be \label{finalequations} y^I+2\lambda y^I y^2={M\over 4\pi r^3}P'\left(\pi^2(0)\right)\pi^I(0)\ .\ee

Dividing these equations by each other, we obtain the relations
\be {{\rm d}\pi^I/{\rm d}r\over {\rm d}\pi^J/{\rm d}r}={\pi^I(0)\over \pi^J(0)}\ ,\ee
which, when integrated from the origin, gives
\be  {\pi^I(r)\over \pi^J(r)}={\pi^I(0)\over \pi^J(0)}\ .\ee
The various components of the solution are therefore always proportional to each other. Thus, by a global $SO(N)$ rotation, we can rotate the solution into one direction in field space, say the $I=1$ direction, so that the solution takes the form $\pi^1\equiv \pi$ and $\pi^I=0$ for $I\not= 1$.  This model therefore exhibits a kind of spontaneous symmetry breaking of the internal $SO(N)$ symmetry, since any non-trivial solution must pick a direction in field space.

Equation~(\ref{finalequations}) now takes the form
\be 
y+2\lambda y^3={M\over 4\pi r^3}P'\left(\pi^2(0)\right)\pi(0)\ .
\label{otheryeqn}
\ee
As $r$ ranges from zero to infinity, the left hand side is monotonic, and is positive or negative depending on the sign of $P'\left(\pi^2(0)\right)\pi(0)$.  For there to be a continuous solution for $y$ as a function of $r$, the left hand side must be invertible when it is positive (negative).  For a solution to exist, this requires (for non-trivial $\lambda$)
\be\lambda>0\ .\ee  
Thus $y$ is also positive (negative), is monotonic with $r$, and ranges from zero to (negative) infinity as $r$ ranges from infinity to zero.  This in turn implies that ${\rm d}\pi/{\rm d}r$ does not cross zero, and hence $\pi$ is monotonic.  

Equation~(\ref{otheryeqn}) yields a solution for $y$, and hence ${\rm d}\pi/{\rm d}r$, as a function of $r$ and the parameters of the theory.  Integrated from $r=0$ to infinity, this will give a relation between $\pi(0)$ and the asymptotic value of the field $\pi(\infty)$.  The asymptotic field value is essentially a modulus of the theory --- it will be set by whatever cosmological expectation value is present. It is a physically meaningful parameter as it affects the coupling to the source by determining $\pi(0)$.

\begin{figure}[h!]
\begin{center}
\includegraphics[height=2.5in]{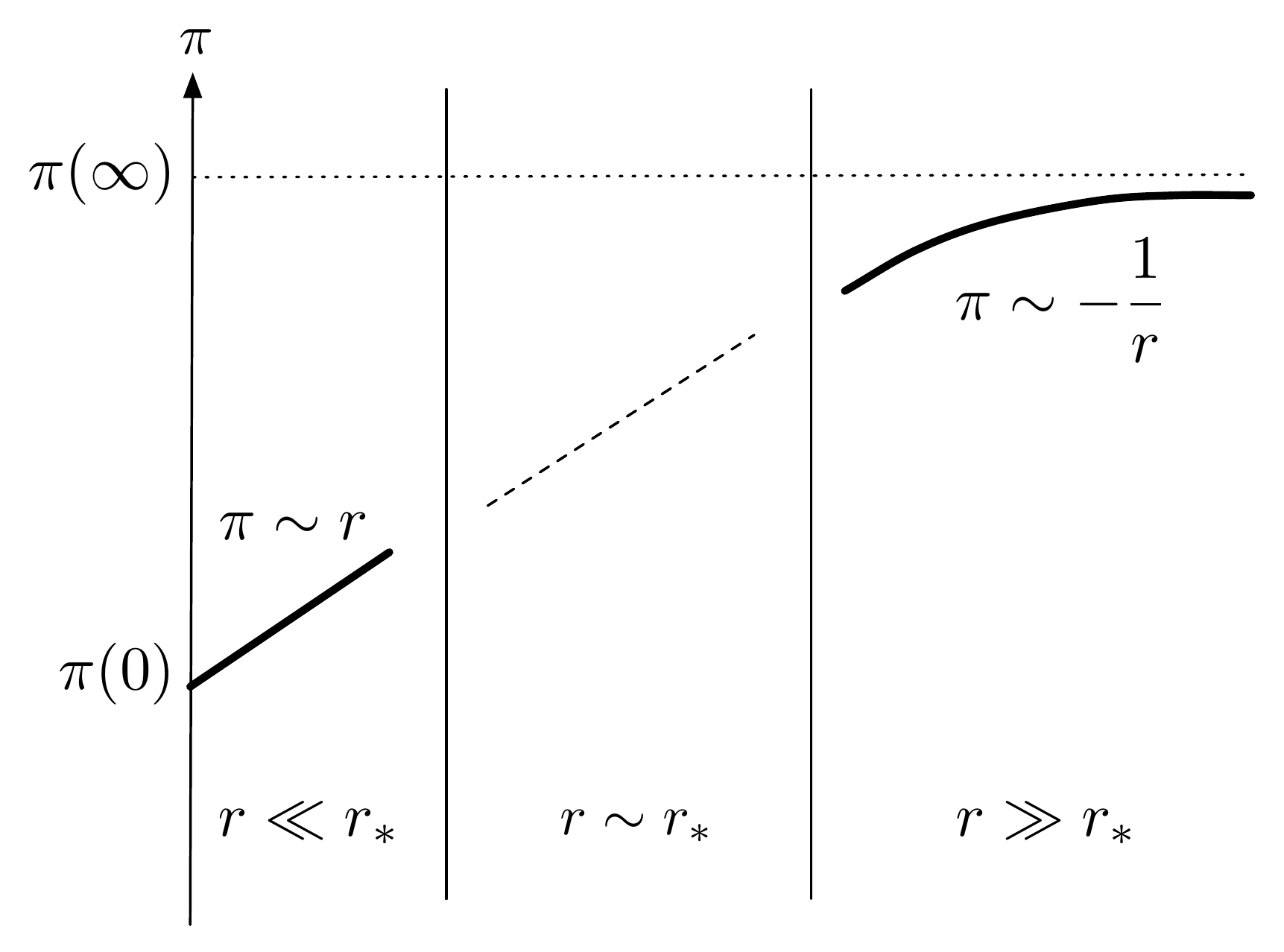}
\caption{Schematic sketch of the solution for $\pi(r)$.}
\label{figure}
\end{center}
\end{figure}

Near the source, where the non-linear term dominates, the solution is linear in $r$, 
\be 
\pi_{r\ll r_\ast}(r) \sim \pi(0)+\left[{M\over 8\pi \lambda}P'\left(\pi^2(0)\right)\pi(0)\right]^{1/3}r \ ,
\ee
whereas far from the source, where the linear term dominates, the solution goes like $1/r$, 
\be 
\pi_{r\gg r_\ast}(r) \sim \pi(\infty)-{M\over 4\pi}P'\left(\pi^2(0)\right)\pi(0){1\over r} \ ,
\ee
where the transition between these regimes occurs at the radius 
\be 
r_\ast\sim\left(\lambda M^2 \left[P'\left(\pi^2(0)\right)\pi(0)\right]^2\right)^{1/6}\ .
\label{vainshteinradius}
\ee
Note that this crossover radius, and hence the distance at which non-linearities become important, depends on the modulus $\pi(0)$.  The equation of motion for
$\pi(r)$ is readily solved numerically, and the solution obtained is plotted schematically in Fig.~\ref{figure}.

\section{Perturbations: Stability and subluminality\label{s:pert}}
While the existence of static, spherically-symmetric configurations is encouraging, there are, of course, other
important checks that our solution must pass to be physically viable. Specifically, following~\cite{Nicolis:2008in}, we must study the stability of these spherically symmetric solutions and to determine the speed at which fluctuations propagate, since superluminal propagation can be an obstacle to finding an ultraviolet completion of the effective theory~\cite{Adams:2006sv}. 

We expand the field in perturbations around the background solution $\pi^I_0$, 
\be \pi^I=\pi^I_0+\delta\pi^I\ .\ee
Away from the source, the linearized equations of motion for the perturbations are of the form
\be
-K^{t}_I(r) \partial_t^2 \delta \pi^I + {1\over r^2} \partial_r \left( r^2 K^{r}_I(r)  \partial_r \delta \pi^I \right)
+ K^{\Omega}_I(r) \partial_\Omega^2 \delta \pi^I = 0\ ,
\ee
where the coefficients $K^{t}_I(r)$, $K^{r}_I(r)$ and $K^{\Omega}_I(r)$ depend on $r$ through the background field $\pi^I(r)$.  We find
\bea
\nonumber
K^t_1&=&{1\over 3r^2}{{\rm d}\over {\rm d}r}\left[r^3\left(1+18\lambda y^2\right)\right]\ ;\\
\nonumber
K^r_1&=&1+6\lambda y^2 \ ;\\
\nonumber
K^\Omega_1&=&{1\over 2r}{{\rm d}\over {\rm d}r}\left[r^2\left(1+6\lambda y^2\right)\right]\ ;\\
\nonumber
K^t_{I\neq 1}&=&{1\over 3r^2}{{\rm d}\over {\rm d}r}\left[r^3\left(1+6\lambda y^2\right)\right]\ ;\\
\nonumber
K^r_{I\neq 1}&=&1+2\lambda y^2 \ ; \\
K^\Omega_{I\neq 1}&=&{1\over 2r}{{\rm d}\over {\rm d}r}\left[r^2\left(1+2\lambda y^2\right)\right] \ .
\label{K1}
\eea

Applying the implicit function theorem to the function $F(y,r)=y+2\lambda y^3-{M\over 4\pi r^3}P'\left(\pi^2(0)\right)\pi(0)=0$, we have 
\be {{\rm d}y\over {\rm d}r}=-{\partial_r F\over \partial_y F}=-{3\over r}{y+2\lambda y^3\over 1+6\lambda y^2}\ .\ee
This allows us to eliminate ${{\rm d}y/{\rm d}r}$ from~(\ref{K1}):
\bea
\nonumber
K^t_1&=&{\left(1-6\lambda y^2\right)^2\over 1+6\lambda y^2}\ ; \\ 
\nonumber
K^r_1&=&1+6\lambda y^2\ ;\\
\nonumber
K^\Omega_1&=&{1-6\lambda y^2\over 1+6\lambda y^2} \ ;\\
\nonumber
K^t_{I\neq 1}&=&{1+12\lambda^2 y^4\over 1+6\lambda y^2}\ ; \\
\nonumber
K^r_{I\neq 1}&=&1+2\lambda y^2 \ ; \\
K^\Omega_{I\neq 1}&=&{1+2\lambda y^2\over 1+6\lambda y^2}\ .
\label{K2}
\eea 

Stability of the spherically symmetric background solutions against small perturbations requires 
$K>0$ for all $K$'s. The $I\neq 1$ directions in field space are stable, but the $\pi^1$ direction exhibits a gradient instability sufficiently
close to the source along the angular directions. In other words, $K^\Omega_1 < 0$ near the source. Therefore, localized perturbations can be found near the source that lower the energy of the solution through their gradients. This instability plagues very short-wavelength fluctuations, right down to the UV cutoff, so decay rates are dominated by the shortest distances in the theory and cannot be reliably computed within the effective theory.

Equations~(\ref{K2}) also allow us to compute the speeds of propagation of our small perturbations, in both the radial and angular directions. These are given by 
\bea
\nonumber
 (c^2)^r_{1} &=& { K^r_1\over  K^t_1}= \left({1+6\lambda y^2}\over{ 1-6\lambda y^2}\right)^2\ ;\\
\nonumber
 (c^2)^\Omega_{1} &=& { K^\Omega_1\over  K^t_1}={1\over 1-6\lambda y^2}\ ; \\
\nonumber
 (c^2)^r_{I\neq 1} &=& { K^r_{I\neq 1}\over  K^t_{I\neq 1}}= {\left(1+2\lambda y^2\right)\left(1+6\lambda y^2\right)\over 1+12\lambda^2 y^4}\ ;\\
 (c^2)^\Omega_{I\neq 1} &=& { K^\Omega_{I\neq 1}\over  K^t_{I\neq 1}} = {1+2\lambda y^2\over 1+12\lambda^2 y^4}\ .
\eea
Note that $(c^2)^r_{1}>1$, and hence these perturbations always propagate superluminally.  The same is true of $(c^2)^\Omega_{1}$, in regions where these perturbations are stable. The speed $(c^2)^r_{I\neq 1}$ is always superluminal, and $(c^2)^\Omega_{I\neq 1}$ is always subluminal.  Whether superluminal propagation of signals is problematic for a low-energy effective theory is still an arguable issue, but it seems that at the least it may preclude the possibility of embedding the theory in a local, Lorentz-invariant UV completion~\cite{Adams:2006sv}.

\subsection{Other constraints\label{constraints}}

It is interesting to note in passing that if a mechanism exists to tame the instabilities we have identified, then precision tests of gravity
within the solar system already place useful constraints on multi-galileon theories.  
The galileon is screened at radii below the Vainshtein radius $r_*$, given by equation~(\ref{vainshteinradius}), restoring the behavior of general relativity. 
Requiring the solar system to be screened to $r \sim 10^{16}\ {\rm m}$ thus yields a constraint on $\lambda$ and $\pi(0)$.
However, lunar laser ranging data constrain the departure from the gravitational potential predicted by GR to satisfy ${\delta\Phi \over \Phi} < 2.4 \times 10^{-11}$ (at radius $r = 3.84 \times 10^{10}$ cm), and we may translate this into a constraint on a different combination of $\lambda$ and $\pi(0)$

For example, consider the choice of $P(X)\mpl\sim\sqrt{\pi^I\pi_I}$, giving a linear coupling between the radial $\pi$ field and matter. In the interesting case when the constraints are saturated, and detection of an effect is therefore imminent, the relevant constraint simply becomes
\be
\frac{1}{\lambda^{1/6}} \lesssim 10^{-9}{\ \rm eV} \ .
\ee
Note that this is an extremely low cutoff for the effective theory, as is also found in the DGP model.
\section{Discussion\label{s:discussion}}
We have derived spherically-symmetric solutions in an $SO(N)$ multi-galileon theory with general, non-derivative couplings to matter.
These solutions exhibit a Vainshtein screening effect, characteristic of galileon models. However, a study of the behavior of fluctuations
around these solutions shows that one of the fields has imaginary sound speed along the angular directions, signaling an instability to
anisotropic modes of arbitrarily short wavelength. Moreover fluctuations inevitably propagate superluminally. 

These results raise serious concerns about the phenomenological viability of $SO(N)$ multi-galileon theories.  (Of course, this does not preclude their effectiveness in early universe physics~\cite{Burrage:2010cu,genesis}, for instance during inflation, as long as they become massive or decouple before the present epoch.)  A key input in
our analysis is the restriction to non-derivative coupling to matter. The main lesson to be drawn is that more general, derivative couplings are
necessary. For instance, the lowest-dimensional coupling invariant under the galilean and internal rotation symmetries is
$\sim\partial_{\mu}\pi^I\partial_{\nu}\pi_I T^{\mu\nu}$. This coupling in fact naturally arises in the higher-codimension brane
picture~\cite{deRham:2010eu}. As mentioned earlier, the galileon fields are oblivious to static, spherically-symmetric sources in this case; thus exhibiting
a screening mechanism. However, they will be excited by orbital motion, and we leave a study of the phenomenological implication of this coupling to future work. 

Our analysis also highlights a distinct advantage to explicitly breaking the symmetry~(\ref{multisymmetry}), for example through the introduction of a sequence of regulating branes of different co-dimensions, as in the cascading gravity case~\cite{cascade1,cascade2,Agarwal:2009gy}. The explicit breaking of $SO(N)$ symmetry allows for more general terms in the action, which can lead to a healthier phenomenology~\cite{Padilla:2010tj}.

Finally, should a creative cure for our instabilities be found, then we have demonstrated that precision solar system tests of gravity set
interesting constraints on multi-galileon theories.

\acknowledgments
 This work is supported in part by NASA ATP grant NNX08AH27G, NSF grant PHY-0930521, and by Department of Energy grant DE-FG05-95ER40893-A020. M.T. is also supported by the Fay R. and Eugene L. Langberg chair. The work of K.H. and J.K. is supported in part by funds from the University of Pennsylvania.


\end{document}